\documentclass[floatfix,
 reprint,
superscriptaddress,
 amsmath,amssymb,
 aps,prl,
]{revtex4-2}

\usepackage{amsmath}
\usepackage{textcomp}
\usepackage{physics}
\usepackage{graphicx}
\usepackage{dcolumn}
\usepackage{bm}
\usepackage[unicode=true, colorlinks=true]{hyperref}
\usepackage[mathlines]{lineno}

\begin{document}

\title{Strongly Contracted N-Electron Valence State Perturbation Theory Using Reduced Density Matrices from a Quantum Computer}

\author{Michal Krompiec}
\email{michal.krompiec@quantinuum.com}
\author{David Mu{\~n}oz Ramo}%
\affiliation{%
 Quantinuum, Partnership House, Carlisle Place, London SW1P 1BX, United Kingdom
}%

\date{\today}

\begin{abstract}
We introduce QRDM-NEVPT2: a hybrid quantum-classical implementation of strongly-contracted N-electron Valence State $2^{nd}$-order Perturbation Theory (SC-NEVPT2), in which the Complete Active Space Configuration Interaction (CASCI) step, capturing static correlation effects, is replaced by a simulation performed on a quantum computer. Subsequently, n-particle Reduced Density Matrices (n-RDMs) measured on a quantum device are used directly in a classical SC-NEVPT2 calculation, which recovers remaining dynamic electron correlation effects approximately. We also discuss the use of the cumulant expansion to approximate the whole 4-RDM matrix or only its zeros. In addition to noiseless state-vector quantum simulations, we demonstrate, for the first time, a hybrid quantum-classical multi-reference perturbation theory calculation, with the quantum part performed on a quantum computer. 

\end{abstract}

\maketitle


\emph{Introduction --} Hamiltonian simulation, the central problem of computational quantum chemistry, is exceptionally well-defined but has long been known to be, as Dirac put it, ``too complex to be solved'' \cite{DiracQuote} in the general case.  Indeed, the exact solution (in a finite basis) of the molecular electronic structure problem, i.e. Full Configuration Interaction (FCI), scales combinatorially with basis size. Hence, the application of FCI is limited to small systems, even in its stochastic approximate implementations \cite{doi:10.1021/acs.jpclett.0c03225}. Quantum computing has recently emerged as a promising approach for large-scale accurate electronic structure calculations, due to exponential (or near-exponential) speedup of certain quantum algorithms, such as Quantum Phase Estimation (QPE) compared to classical solutions such as FCI \cite{AspuruGuzik2005,cao2019quantum, mcardle2020quantum,Troyer_Reiher_2021_rev,motta2021emerging, Liu2022}. 

\emph{Towards quantum multi-reference calculations --} While the motivation for the development of quantum algorithms for computational chemistry seems to come from the desire for an exact solution of the chemical Hamiltonian simulation problem (i.e. a quantum replacement for FCI), exact diagonalization of the whole Hamiltonian of a chemical system is in practice hardly ever pursued, or even needed. The \textit{special case} of ground-state energies of ``single-reference'' molecular systems (i.e. exhibiting mainly weak correlations and having one dominant configuration in the Configuration Interaction expansion) can be very accurately calculated with Coupled Cluster methods, such as CCSD(T), which scales with basis size as $\mathcal{O}(N^7)$  in the canonical implementation and only $\mathcal{O}(N)$ in the DLPNO approximation \cite{NeeseDLPNOCCSDT}. The remaining strongly correlated electronic systems can usually be described by \textit{multi-reference} or \textit{multi-configurational} methods, where interactions within only a subspace of the Hilbert space are calculated with high accuracy, while interactions with remaining orbitals are treated only approximately. Thus, the orbitals are divided into two disjoint sets: the active orbitals and the inactive (core and virtual) orbitals. A model Hamiltonian is defined in the reference (or model) subspace defined by Slater determinants generated by permutations of active orbitals and accounts chiefly for the static electron correlation. An expansion in which the model space includes all possible distributions of electrons in the selected (active) orbitals is called complete active space (CAS) \cite{mcscf_ms_gordon1998,doi:10.1063/1.5039496}. In the first step of a multi-reference calculation, static correlation effects are introduced with a variationally optimized reference wave function: 
\begin{equation} 
\ket{\Psi_0} = \sum_{\mu=1}^{d} c_{\mu}\ket{\Psi_{\mu}} 
\end{equation}

Subsequently, dynamical correlation effects are introduced via a wave operator $\Omega$ acting on $\ket{\Psi_0}$ \cite{doi:10.1063/1.5039496}. The action of $\Omega$ is often approximated at a lower level of theory, for example perturbation theory truncated at $2^{nd}$ order, like in the popular CASPT2 \cite{Pulay2011} and NEVPT2 \cite{Angeli2001,Angeli2002} methods. For large active spaces, the cost of CASPT2 and NEVPT2 calculations is dominated by the solution of the CAS problem which scales exponentially with the size of the active space due to an exponential scaling of the CI basis. By mapping the electronic occupation number vectors of length N to a quantum register (i.e. to qubits), it becomes possible to express this enormous CI basis via the $2^N$ basis states of N qubits \cite{mcardle2020quantum}; the same mapping applied to the electronic Hamiltonian yields the corresponding qubit Hamiltonian. Determination of its the ground state via e.g. QPE requires a number of steps only polynomial in N \cite{mcardle2020quantum}, suggesting an exponential speedup with respect to the classical approach. When the cost of preparing the initial state with non-negligible overlap with the ground state is factored in, the quantum speedup is expected to be less than exponential but still potentially very large \cite{garnet_exp_advantage}.
Hence, replacing the CAS-CI component of the calculation with an efficient quantum algorithm would allow application of these techniques to very large active spaces, thereby extending their applicability to extended, complex chemical systems and eliminating the need for selection of active orbitals. 

NEVPT2(VQE,QSE) is a hybrid quantum-classical implementation of uncontracted NEVPT2 recently published by Tammaro et al. \cite{Tammaro2022}, in which the CAS calculation is replaced by the Variational Quantum Eigensolver (VQE) \cite{peruzzo2014variational, https://doi.org/10.48550/arxiv.2111.05176}. After state preparation with VQE, n-particle Reduced Density Matrices (RDMs) up to $n = 4$ are measured and Quantum Subspace Expansion (QSE) \cite{qse_PhysRevA.95.042308} algorithm with single and double excitations as the expansion operators is applied to determine all eigenvectors and eigenvalues of the active space (Dyall) Hamiltonian (in the subspace defined by the chosen expansion operators). Finding all excited states via QSE, is expensive and will likely become intractable for large active spaces, potentially cancelling out any quantum advantage of the state preparation step. Therefore, for complex multireference systems (for which quantum advantage is expected in the CAS part of this workflow), NEVPT2(VQE,QSE) will likely be prohibitively expensive and, if a limited rank  of expansion operators in QSE is used, inaccurate.

\emph{QRDM-NEVPT2 --} In the present work, we introduce \emph{QRDM-NEVPT2}: a hybrid quantum-classical flavor of SC-NEVPT2 \cite{Angeli2002,Guo2016} implemented in the quantum chemistry package InQuanto \footnote{https://www.quantinuum.com/products/inquanto}. Following Tammaro et al. \cite{Tammaro2022}, we apply VQE in place of the CAS-CI step, but in contrast to their method, we do not rely on QSE but use RDMs computed via measurement of expectation values of RDM operators after state preparation with VQE. 

\begin{figure*}
\includegraphics[width=14cm]{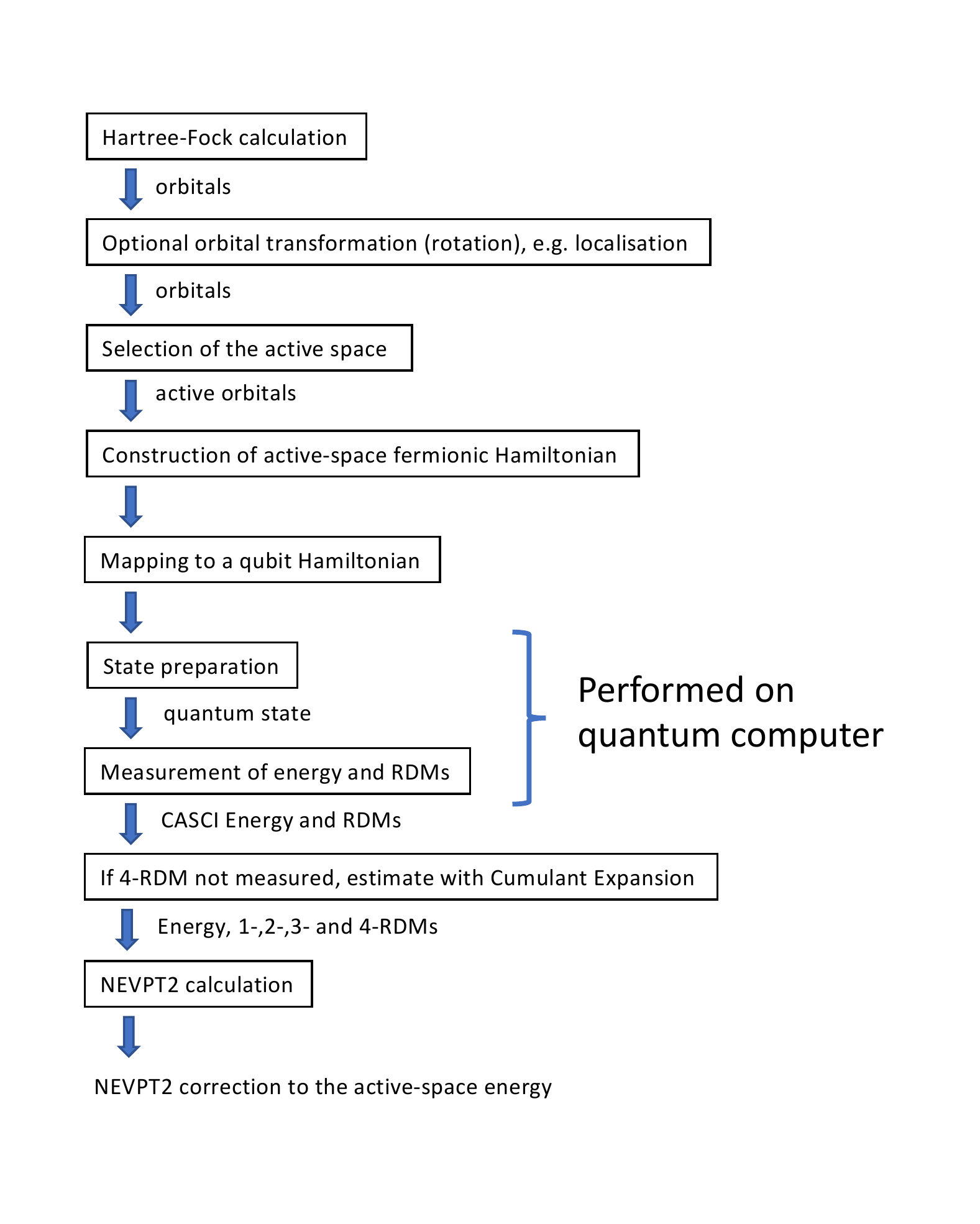}
\caption{\label{fig:vqe-nevpt2}Flowchart of the QRDM-NEVPT2 method. The `State preparation', in the present implementation, is VQE.}
\end{figure*}

Our workflow (see Fig.~\ref{fig:vqe-nevpt2}) 
starts with the usual classical bootstrapping of a quantum simulation \cite{https://doi.org/10.48550/arxiv.2111.05176}: a mean-field calculation followed by an optional orbital transformation (e.g. localization), selection of the active space, construction of a fermionic $2^{nd}$-quantized Hamiltonian and Jordan-Wigner mapping \cite{wigner1928paulische, AspuruGuzik2005} it into a qubit Hamiltonian. The state-preparation step in our implementation consists of VQE \cite{peruzzo2014variational, https://doi.org/10.48550/arxiv.2111.05176}, which also yields the expectation value of the active space Hamiltonian.  Subsequently, we use the Operator Averaging \cite{peruzzo2014variational} method to jointly measure the expectation values of the matrix elements of spin-traced 1-, 2- and 3-RDM operators. In order to reduce the number of required measurements, we partition the Pauli words defining the RDM operators into mutually commuting sets, each set corresponding to one measurement circuit \cite{cowtan2020generic}. We make use of Z$_2$ symmetries of the qubit Hamiltonian to mitigate errors via Partition Measurement Symmetry Verification \cite{yamamoto2021} and to reduce quantum resources via qubit tapering \cite{tapering}. We note that RDM operator matrix elements which violate any of the Z$_2$ symmetries of the qubit Hamiltonian vanish and therefore do not need to be measured. For systems with more than 3 active electrons (i.e. where the 4-RDM does not vanish), we estimate the spin-traced 4-RDM using the cumulant approximation formula (see below). Once the RDMs are determined, we compute the NEVPT2 energy using a modification of Guo's implementation \cite{Guo2016} of SC-NEVPT2 \cite{Angeli2002}, where we replaced RDMs calculated from CAS-CI or DMRG wave functions with those obtained from our algorithm. 

\emph{The cumulant approximation of 4-RDM --} The spin-traced excitation operators are defined by
\begin{subequations}
\begin{equation}
    \hat{E}^{p}_{q} = \hat{a}^{\dag}_{p\uparrow}\hat{a}_{q\uparrow}+\hat{a}^{\dag}_{p\downarrow}\hat{a}_{q\downarrow}
\end{equation}
\begin{eqnarray}
    \hat{E}^{pr}_{qs} = 
    \hat{a}^{\dag}_{p\uparrow}\hat{a}^{\dag}_{r\uparrow}\hat{a}_{s\uparrow}\hat{a}_{q\uparrow}
   +\hat{a}^{\dag}_{p\uparrow}\hat{a}^{\dag}_{r\downarrow}\hat{a}_{s\downarrow}\hat{a}_{q\uparrow}\nonumber\\
   +\hat{a}^{\dag}_{p\downarrow}\hat{a}^{\dag}_{r\uparrow}\hat{a}_{s\uparrow}\hat{a}_{q\downarrow}
   +\hat{a}^{\dag}_{p\downarrow}\hat{a}^{\dag}_{r\downarrow}\hat{a}_{s\downarrow}\hat{a}_{q\downarrow}, 
\end{eqnarray}
\end{subequations}
etc. The spin-traced 4-RDM ($\Gamma^{prtv}_{qsuw}$) is defined as
\begin{equation} 
\Gamma^{prtv}_{qsuw} = \bra{\Psi_0} \hat{E}^{prtv}_{qsuw} \ket{\Psi_0} 
\end{equation}

PySCF's \cite{Guo2016,pyscf} and Orca's \cite{neese_cu4} implementations of NEVPT2 make use of a related tensor instead, dubbed 4-particle Pre-Density Matrix (4-PDM, $\gamma^{prtv}_{qsuw}$) by Guo et al. \cite{neese_cu4}. PDM is defined with the same creation and annihilation operators as the corresponding RDM, but applied in a different order:

\begin{equation} 
\gamma^{prtv}_{qsuw} = \bra{\Psi_0} \hat{E}^{p}_{q}\hat{E}^{r}_{s}\hat{E}^{t}_{u}\hat{E}^{v}_{w} \ket{\Psi_0} 
\end{equation}
Hence, the 4-RDM can be computed easily from 4-PDM and lower RDMs, via normal-ordering of the operators in 4-PDM \cite{neese_cu4}. 

4-RDM and 4-PDM are tensors consisting of $N^8$ elements, where N is the number of spatial orbitals, hence their measurement on a quantum device would be very expensive. Approximation of higher-order RDMs via the cumulant expansion (i.e. setting the highest-order cumulant to 0) is well known \cite{valdemoro1993,harris2002cumulant}, but an analogous expression for spin-traced RDMs has been introduced by Kutzelnigg et al. \cite{Kutzelnigg2010} only in 2010. Their formula connecting the spin-traced 4-particle density cumulant $\Lambda^{P_1P_2P_3P_4}_{Q_1Q_2Q_3Q_4}$ with the spin-traced 4-particle RDM  $\Gamma^{P_1P_2P_3P_4}_{Q_1Q_2Q_3Q_4}$ and lower-order terms reads: 
\begin{eqnarray}
\Lambda^{P_1P_2P_3P_4}_{Q_1Q_2Q_3Q_4} = \Gamma^{P_1P_2P_3P_4}_{Q_1Q_2Q_3Q_4}(1) - 
\Gamma^{P_1}_{Q_1}\Lambda^{P_2P_3P_4}_{Q_2Q_3Q_4}(4) \nonumber\\
- \Gamma^{P_1}_{Q_1}\Gamma^{P_2}_{Q_2}\Lambda^{P_3P_4}_{Q_3Q_4}(6) 
- \Lambda^{P_1P_2}_{Q_1Q_2}\Lambda^{P_3P_4}_{Q_3Q_4}(3) \nonumber\\
+\frac{1}{2}\Bigl\{\Gamma^{P_1}_{Q_1}\Lambda^{P_2P_3P_4}_{Q_1Q_3Q_4}(12)
+ \Gamma^{P_1}_{Q_2}\Gamma^{P_2}_{Q_1}\Lambda^{P_3P_4}_{Q_3Q_4}(6) \nonumber\\
+ \Gamma^{P_1}_{Q_1}\Gamma^{P_2}_{Q_3}\Lambda^{P_3P_4}_{Q_2Q_4}(24) +
\Lambda^{P_1P_2}_{Q_1Q_3}\Lambda^{P_3P_4}_{Q_2Q_4}(12)\Bigl\}\nonumber\\
-\frac{1}{4}\Bigl\{\Gamma^{P_1}_{Q_2}\Gamma^{P_2}_{Q_3}\Lambda^{P_3P_4}_{Q_1Q_4}(24)
+\Gamma^{P_1}_{Q_3}\Gamma^{P_2}_{Q_4}\Lambda^{P_3P_4}_{Q_1Q_2}(12)\Bigl\}\nonumber\\
-\frac{1}{12}\Bigl\{\left(\Lambda^{P_1P_2}_{Q_3Q_4}-\Lambda^{P_1P_2}_{Q_4Q_3}\right)
\left(\Lambda^{P_3P_4}_{Q_1Q_2}-\Lambda^{P_3P_4}_{Q_2Q_1}\right)\nonumber\\
+3\left(\Lambda^{P_1P_2}_{Q_3Q_4}+\Lambda^{P_1P_2}_{Q_4Q_3}\right)
\left(\Lambda^{P_3P_4}_{Q_1Q_2}+\Lambda^{P_3P_4}_{Q_2Q_1}\right)\Bigl\}\nonumber\\
\Gamma^{P_1}_{Q_1}\Gamma^{P_2}_{Q_2}\Gamma^{P_3}_{Q_3}\Gamma^{P_4}_{Q_4}(1)
+\frac{1}{2}\Gamma^{P_1}_{Q_2}\Gamma^{P_2}_{Q_1}\Gamma^{P_3}_{Q_3}\Gamma^{P_4}_{Q_4}(6)\nonumber\\
-\frac{1}{4}\Gamma^{P_1}_{Q_2}\Gamma^{P_2}_{Q_3}\Gamma^{P_3}_{Q_1}\Gamma^{P_4}_{Q_4}(8)
-\frac{1}{4}\Gamma^{P_1}_{Q_3}\Gamma^{P_2}_{Q_4}\Gamma^{P_3}_{Q_1}\Gamma^{P_4}_{Q_2}(3)\nonumber\\
+\frac{1}{8}\Gamma^{P_1}_{Q_2}\Gamma^{P_2}_{Q_3}\Gamma^{P_3}_{Q_4}\Gamma^{P_4}_{Q_1}(6).
\end{eqnarray}
The numbers in parentheses denote numbers of q-permutations of indices (e.g. (3) means that the preceding term is a sum of 3 terms with permuted indices). Approximate $\Gamma^{P_1P_2P_3P_4}_{Q_1Q_2Q_3Q_4}$ is calculated by setting $\Lambda^{P_1P_2P_3P_4}_{Q_1Q_2Q_3Q_4}=0$. In our code, we use Saitow's \cite{Saitow2013} Fortran implementation of Kutzelnigg's formula. 

We note, however, that Zgid et al. found that replacing 3- and 4-RDMs by the respective cumulant approximations in SC-NEVPT2 leads to significant degradation of accuracy and appearance of intruder states \cite{zgid_cu4}. Moreover, the impact of using cumulant approximation to 4-RDM (hereafter called the CU(4) approximation) in Fully Internally Contracted (FIC) NEVPT2 was shown to be limited for simpler cases (such as photoisomerization of stilbene) but profound or even catastrophic for highly-multireference systems, such as stretched N$_2$ or Cr$_2$ \cite{neese_cu4}. We conclude that cumulant approximations should be used with great caution in NEVPT2 as they are expected to fail for highly multi-reference systems. 

\begin{figure*}
\includegraphics[width=8cm]{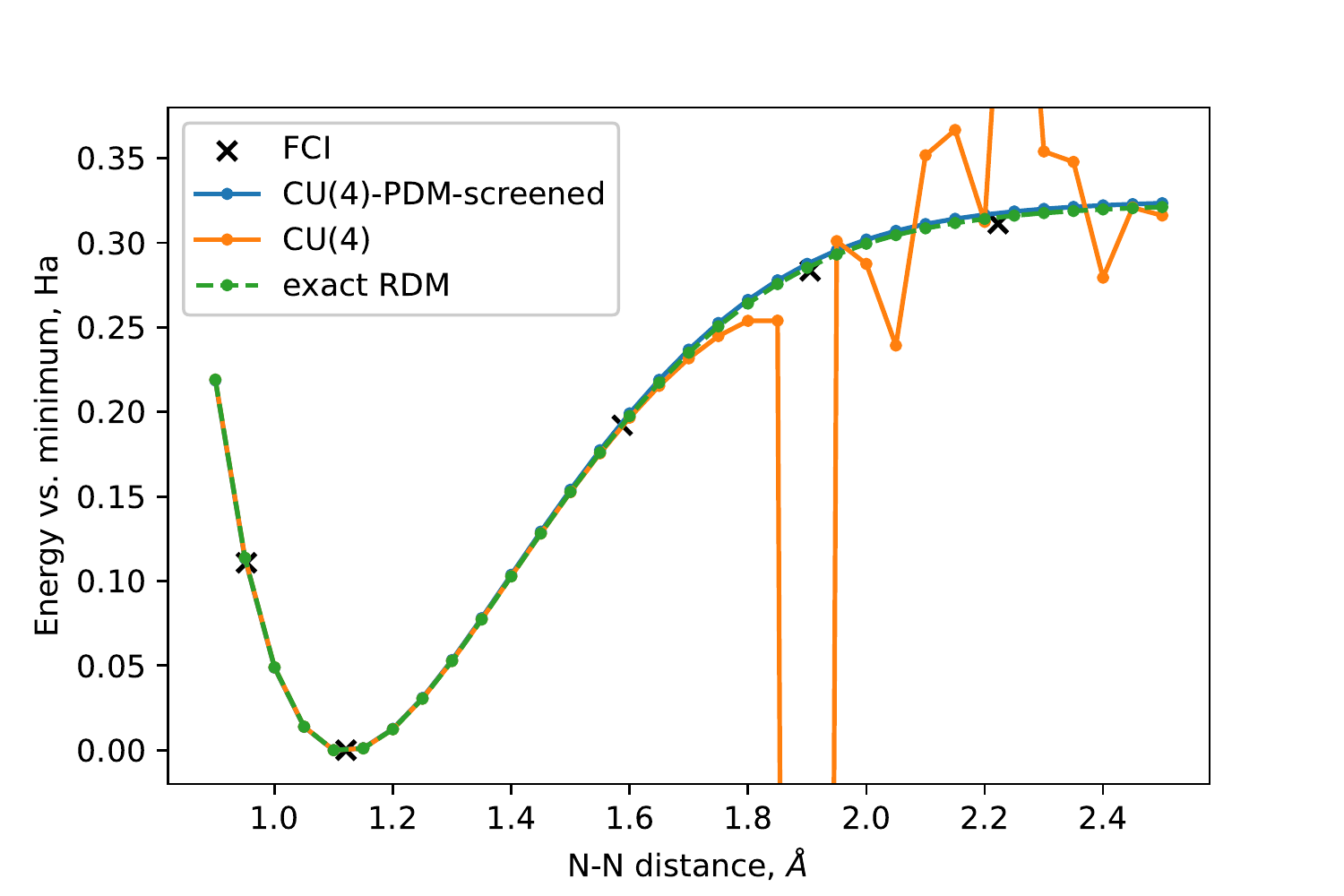} \includegraphics[width=8cm]{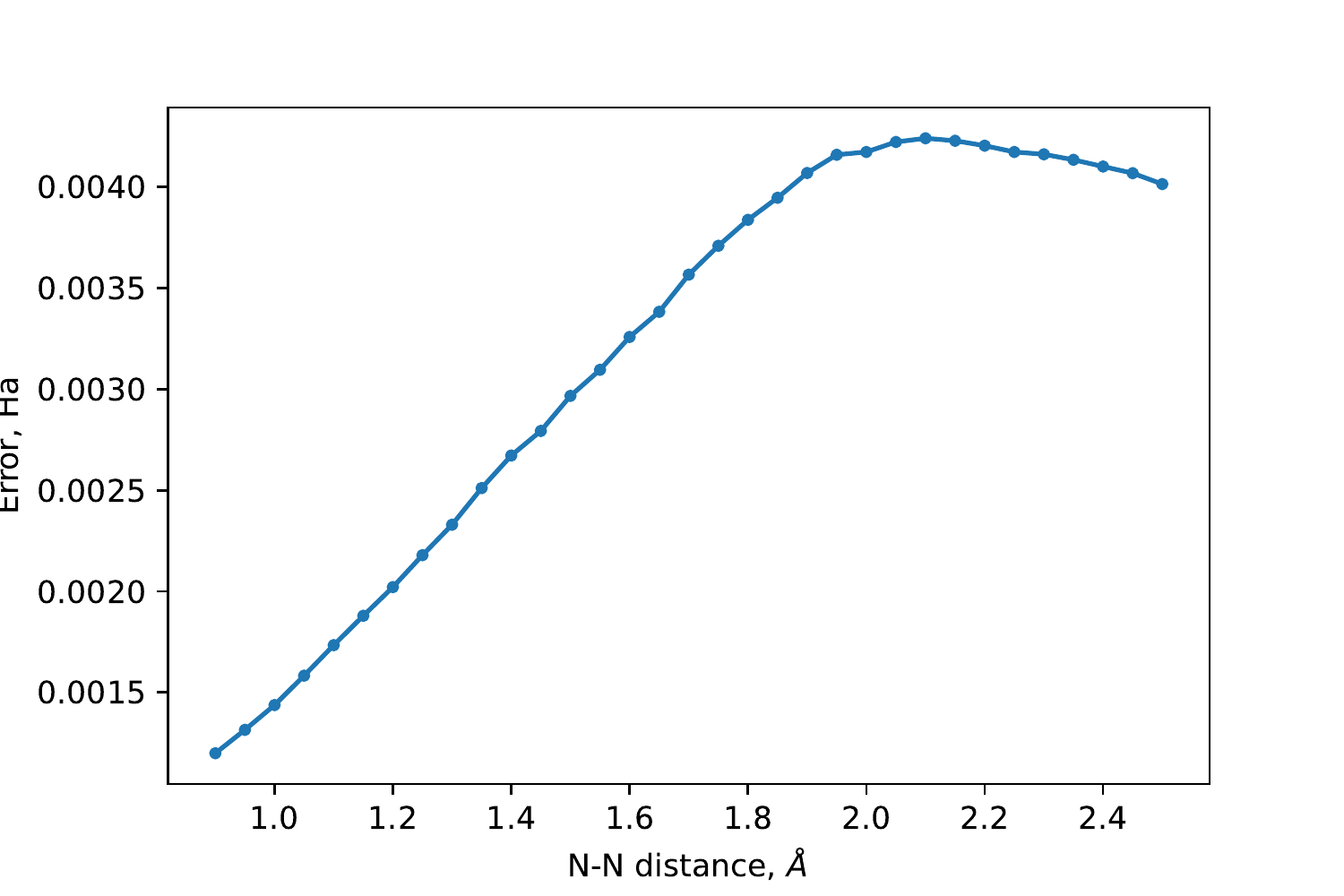}
\caption{\label{fig:cu4_approx_filter} Effects of approximating 4-RDM on the N$_2$ binding energy calculated at CAS(10,8)-SC-NEVPT2/cc-pVDZ level compared to FCI/cc-pVDZ. Left: NEVPT2 dissociation curves obtained from exact and approximated 4-RDM compared with FCI reference. Right: error of NEVPT2 binding energies caused by CU(4)-PDM-screened approximation.}
\end{figure*}

\emph{Novel approximations to 4-RDM and 4-PDM -- } We investigated novel methods for approximation of 4-particle RDM and PDM for use in SC-NEVPT2, using N$_2$ dissociation curve at CAS(10,8)-SC-NEVPT2/cc-pVDZ level as the test case and FCI/cc-pVDZ data \cite{chan2004state} as benchmark \footnote{All classical calculations were carried out with PySCF 2.0.1 \cite{pyscf}}. First, we noticed that the exact 4-RDM matrix is relatively sparse: only about 5\% of matrix elements are non-zero at any point of the dissociation curve. We hypothesise that if the vanishing elements of 4-RDM are predicted correctly in the CU(4) approximation, it suffices to compute exactly (i.e. measure) only the matrix elements which are non-zero in the CU(4) approximated 4-RDM. To test this approximation, which we dubbed \textit{CU(4)-RDM-filtered}, we replaced significantly non-zero matrix elements of CU(4) 4-RDM (i.e. having absolute value greater than $10^{-16}$) by the corresponding elements of exact 4-RDM and used the resultant matrix in SC-NEVPT2 calculation. 
Counter-intuitively, instead of improvement vs. CU(4), large errors in the SC-NEVPT2 energies, up to 2.5 Ha, were observed.  We have further noticed that the 4-PDM matrix is also sparse, with about 12\% density across the dissociation curve. 
We define the \textit{CU(4)-PDM-filtered} approximation by replacing the significantly non-zero elements of 4-PDM computed from CU(4)-approximated 4-RDM with the exact values. With this approximation, we improve significantly upon CU(4) and obtain a smooth dissociation curve with only modest deviation from exact NEVPT2 at larger internuclear separations, see Fig. \ref{fig:cu4_approx_filter}. We expected to observe savings in the measurement budget, due to a reduced number of measured matrix elements. We investigate the cost of approximations to 4-RDM and 4-PDM on a simpler example, for which compilation of the circuits for exact 4-RDM is easier: stretched Li$_2$ (Li-Li distance 6.68 \AA) in cc-pVTZ basis and active space of 4 electrons in 4 or 5 orbitals. As before, we use InQuanto and TKET via pytket \cite{sivarajah2020t}; RDMs are measured jointly via Pauli-word partitioning  \cite{cowtan2020generic}.
As seen in Table~\ref{tab:table1}, direct measurement of all n-RDMs up to $n = 4$ requires only just as many or 16 additional measurement circuits (depending on the size of the active space) over 3-RDM. However, the new approximation does not yield any savings in quantum computer time, because the 4-PDM elements are measured separately from the lower-rank matrices, while direct determination of all RDMs involves joint measurement of all matrices via TKET's measurement reduction scheme. 

\begin{table}[b]
\caption{\label{tab:table1}%
Comparison of number of measurement circuits required to measure RDMs for NEVPT2 calculation on stretched Li$_2$, as a function of approximation to 4-RDM and active space size. 
}
\begin{ruledtabular}
\begin{tabular}{cccc}
\textrm{Active space}&\textrm{Approximation}&\textrm{VQE}&\textrm{RDM}\\
\colrule
(4,4) & None  & 13 & 31 \\
(4,4) & CU(4)  & 13 & 31 \\
(4,4) & CU(4)-PDM-filtered & 13 & 60  \\
(4,5) & None  & 23 & 79 \\
(4,5) & CU(4) & 23 & 63  \\
(4,5) & CU(4)-PDM-filtered & 23 & 132 \\

\end{tabular}
\end{ruledtabular}
\end{table}

\begin{figure*}
\includegraphics[width=8cm]{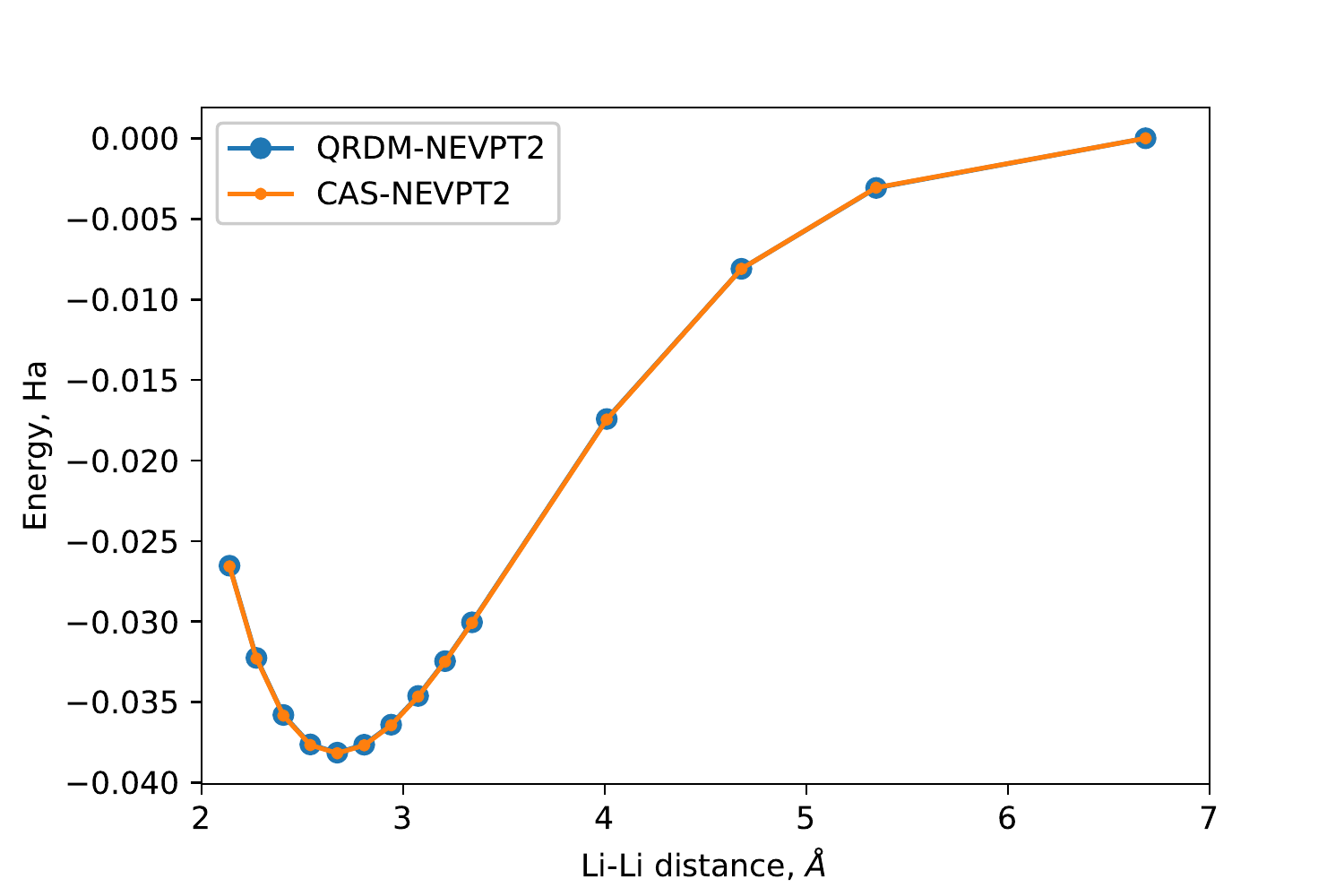} \includegraphics[width=8cm]{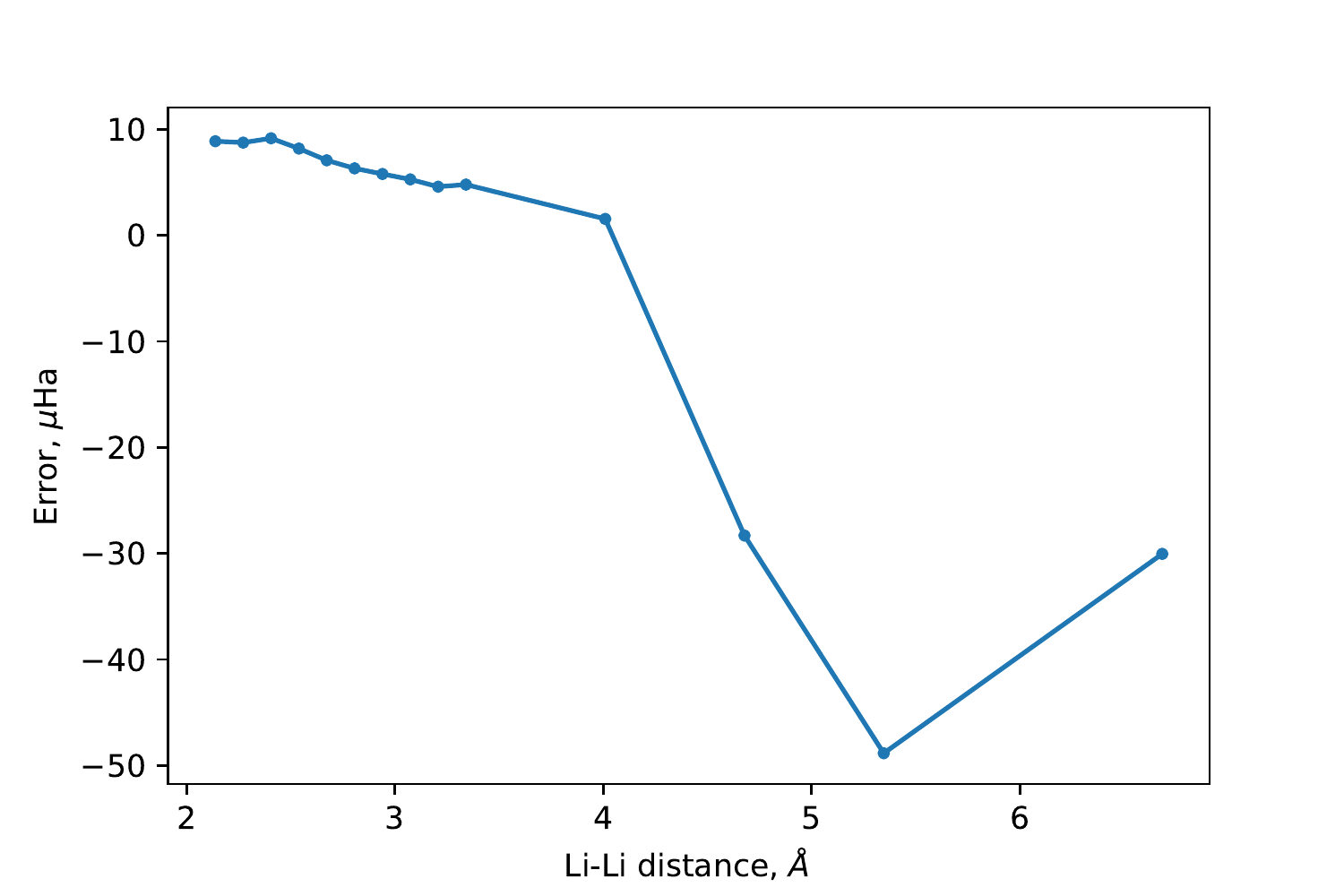} 
\caption{\label{fig:li2} Left: dissociation curve of Li$_2$ calculated with QRDM-NEVPT2 (statevector simulation) compared with classical SC-NEVTP2 (see text). Right: deviation of QRDM-NEVPT2 from classical SC-NEVPT2. }
\end{figure*}

\emph{QRDM-NEVPT2: statevector simulations --} We have applied the QRDM-NEVPT2 workflow to calculation of the dissociation curve of Li$_2$, with an active space of 4 electrons in 6 spatial orbitals (optimized by CASSCF) and using the cc-pVTZ basis set \cite{dunning1989gaussian}. We used our Chemically-Aware variant \cite{chemically-aware} of the UCCSD ansatz \cite{D1CS00932J}. The quantum simulation was performed in InQuanto using the Qulacs statevector simulator \cite{suzuki2021qulacs}. Fig. \ref{fig:li2} shows its excellent agreement with a reference SC-NEVPT2 calculation (up to 50 $\mu$Ha error), despite the use of the CU(4) approximation. 

\begin{figure*}
\includegraphics[width=8cm]{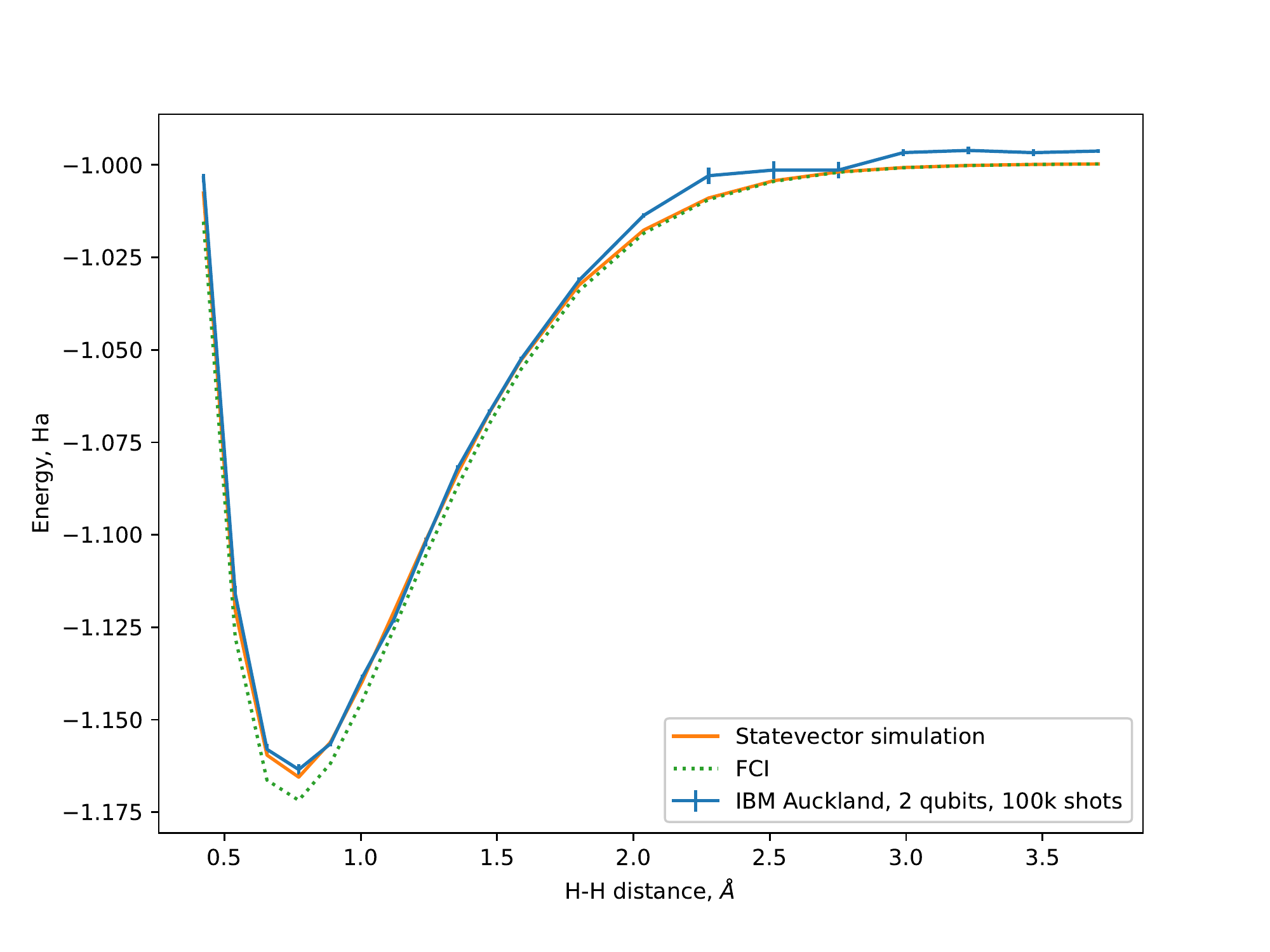} 
\includegraphics[width=8cm]{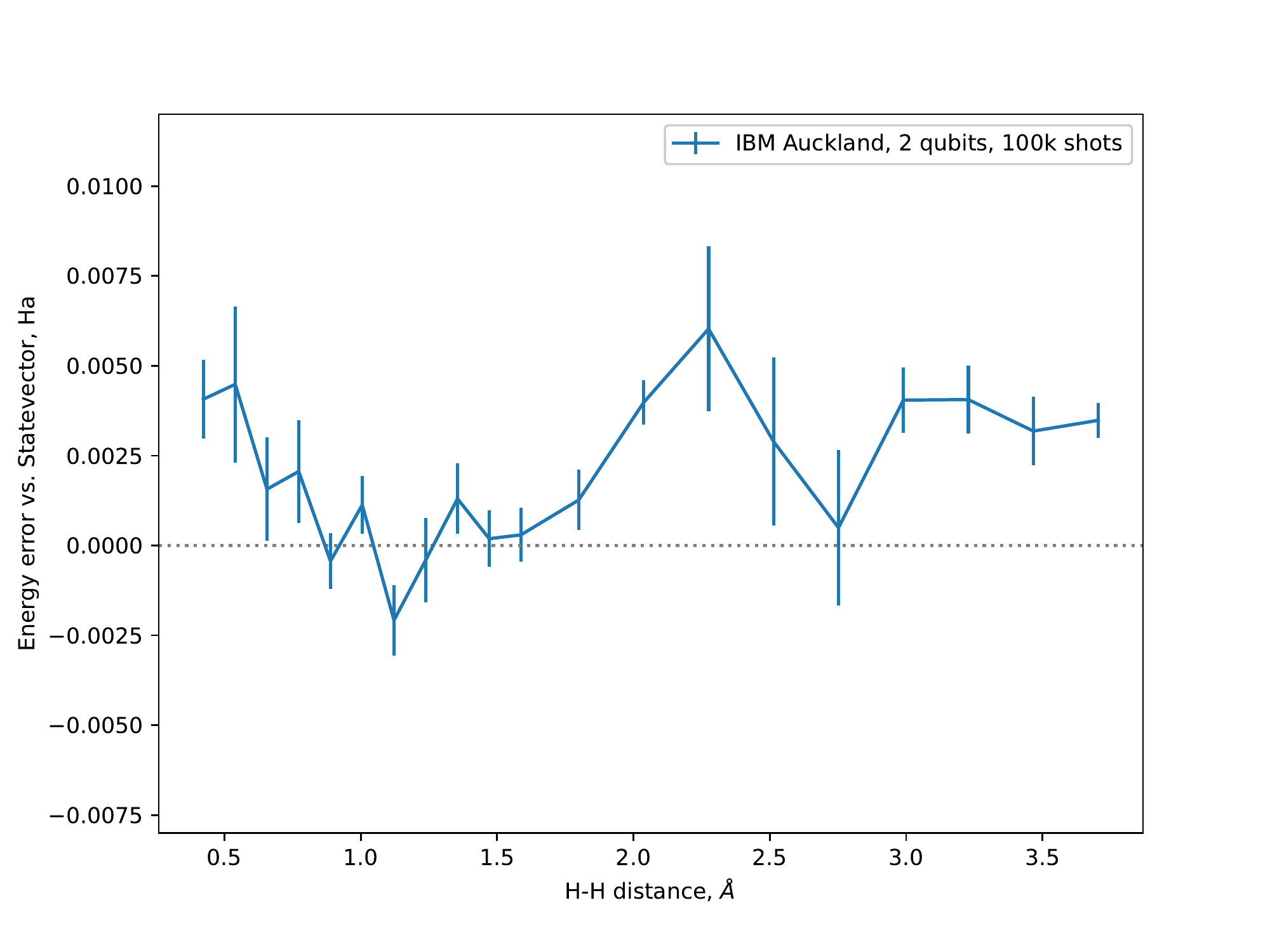} 
\caption{\label{fig:h2} Dissociation curve of H$_2$ calculated with QRDM-NEVPT2. Active space: (2,2), basis set: cc-pVTZ. IBM Auckland data represent means of 100k measurements grouped in 10 batches 10k samples each, error bars show standard errors of the mean. Left: absolute energies, right: deviations from noiseless (statevector) simulation. 
}
\end{figure*}

\emph{QRDM-NEVPT2: hardware and noisy emulator experiments --} To test QRDM-NEVPT2 on today's quantum devices, we have chosen the H$_2$ dissociation curve as a test case. We used the cc-pVTZ basis set and the active space consisted of two electrons in two spatial orbitals (canonical Hartree-Fock orbitals were used). The quantum register was tapered off \cite{tapering} from 4 qubits to 2, using the IZZI and IIZZ symmetry operators of the qubit Hamiltonian. We applied PMSV error mitigation \cite{yamamoto2021}, taking advantage of the ZZ symmetry of the tapered Hamiltonian. Fig. \ref{fig:h2} shows the comparison of QRDM-NEVPT2 simulated in Qulacs \cite{suzuki2021qulacs} statevector simulator to that executed on superconducting transmon IBM Auckland \cite{ibm_services_list} device.

\emph{Conclusions and outlook --} We have introduced a hybrid quantum-classical implementation of SC-NEVPT2, consisting of a combination of variational state preparation and measurement of RDMs on a quantum computer followed by classical calculation of NEVPT2 energy.  We note that for the state-preparation, a non-variational algorithm (e.g. such as Quantum Imaginary Time Evolution \cite{Motta_2019}) could potentially be used instead of VQE and the RDMs could be measured by e.g. quantum overlapping tomography \cite{cotler2020quantum}, thus enabling high accuracy and scalability to large systems on a future Fault-Tolerant quantum computer. We have investigated a novel approximation to 4-RDM, \textit{CU(4)-RDM-filtered} which reduced the number of matrix elements to evaluate but resulted in an overall increase in the number of measurement circuits. 

Our approach ensures that the quantum electronic structure solver, being essentially a drop-in replacement for CAS-CI, is applied only to the \textit{most strongly correlated} sub-problem of the electronic structure problem, while the remaining (mostly weak) correlations are treated at the level of perturbation theory. Thus, QRDM-NEVPT2 mirrors classical multireference methods and is in stark contrast to prevalent practice in prototypical applications of VQE \cite{https://doi.org/10.48550/arxiv.2111.05176} and resource estimation of QPE \cite{qubitization2019, QuNCo_Schrodinger_2020} where interactions outside of the active space remain at Hartree-Fock level, as in classical CASCI and CASSCF methods. We argue that the latter approach is unlikely to be practically useful. Neglect of electron correlation outside of the active space is tantamount to neglecting most of dynamic correlation energy, unless the active space is huge (e.g. covers more than half of the orbitals \cite{Bartlett_FNO_2008}). However, correlating all orbitals would require very large resources just to map the quantum state, e.g. a single benzene molecule in cc-pVQZ basis would need 1020 qubits to encode occupation of all spinorbitals: an order of magnitude more than the 108-qubit active space considered in resource estimation for QPE of FeMoco \cite{qubitization2019}. Despite the inherent limitations of QRDM-NEVPT2, shared by all 2$^{nd}$ order perturbation theories, we believe it may represent the best way of computing dynamic correlation corrections to active-space-type quantum methods. Extending our workflow to higher-level theories such as IC-MR-CC would not be straightforward and would require measurement of even higher-rank RDMs \cite{doi:10.1063/1.5039496,hanauer2013internally}.

\emph{Conflict of interest --} We declare the use of IBM Quantum services for this work. The views expressed are those of the authors, and do not reflect the official policy or position of IBM and Quantinuum. IBM also own shares in Quantinuum.

\begin{acknowledgments}
M.K. is grateful to Nathan Fitzpatrick, Iakov Polyak, David Zsolt Manrique, Irfan Khan, Cono Di Paola, Gabriel Greene-Diniz and Josh Kirsopp for helpful discussions.  The state-vector simulations described in this work were performed on Microsoft Azure Virtual Machines provided by the ``Microsoft for Startups'' program.
\end{acknowledgments}

\bibliography{vqe_nevpt2}
\end{document}